\documentstyle[preprint,revtex]{aps}
\math-with-secnums
\begin{document}
\draft
\begin{title}
\bf{ N-body theory revisited \\
and its extension to the $\pi$NNN--NNN problem $^*$}
\end{title}
\author{G. Cattapan and L. Canton}
\begin{instit}
Istituto Nazionale di Fisica Nucleare, Sezione di Padova  \\
Dipartimento di Fisica, Universit\'a di Padova \\
via Marzolo, 8 I-35131 Padova
\end{instit}

\begin{abstract}
In order to approach the pion--multinucleon problem,
we have found it convenient to
reformulate the general $N$--body theory starting from the fully
unclusterized ({\it i.e.}, $N\leftarrow N$) amplitude.
If we rewrite such an amplitude in terms of new unknowns which
can be later identified as the amplitudes for all the
$(N-1)\leftarrow (N-1)$ cluster processes, and repeat recursively
the procedure up to the treatment of the $2 \leftarrow 2$ cluster processes,
we obtain very naturally the hierarchy of equations which ranges from
the $N$--body fully--disconnected Lippmann--Schwinger equation to the
$N$--body connected--kernel Yakubovski\u\i--Grassberger--Sandhas one.
This revisitation turns out to be very useful when considering the
modifications required in case one of the bodies is a pion and the
remaining are nucleons, with the pion being allowed to disappear and
reappear through the action of a pion--nucleon vertex. In fact, we
obtain a new set of coupled pion--multinucleon equations which allow
a consistent and simultaneous treatment of $\pi$ scattering and absorption.
For the $\pi$NNN system, the kernel of these coupled equations is shown
to be connected after three iterations.
\par
\
\end{abstract}
\newpage
\narrowtext

\section{Introduction}

      $N$--body scattering theory is by now well understood.
We have
a good insight into its basic structural features, and understand
the relations between the various formulations of the multiparticle
collision problem both in the wave--function and in the
transition--operator language \cite{sa78,sa80,cv83f}. Some important
hints have also been gained about how to tackle the formidable
computational problems implied by few--body scattering theory, at least
in the four--body case \cite{note1}. Among the many formulations of
$N$--body scattering theory, the approach developed by Sandhas and
co--workers \cite{ags,gs67,agsp} plays a privileged role. Indeed,
the Alt--Grassberger--Sandhas (AGS)
formulation of scattering theory, while being fully equivalent to the
celebrated Yakubovski\u\i\ \cite{ya67} approach \cite{agsp,kz74}, is
at the same time susceptible of a much clearer physical interpretation
thanks to the employment of a powerful matrix notation.
\par
      The above situation has to be contrasted with what happens
for the $\pi$--multinucleon problem. In such a case, the
possibility of production or absorption of the meson greatly
complicates the theory, since the presence of production/absorption
processes prevents one from using standard few--body connected--kernel
methods, which essentially represent a clever resummation
of non--relativistic multiple--scattering diagrams
(with a {\it fixed}, finite--number of degrees of freedom).
\par
      When dealing with $\pi$--multinucleon systems, one is faced
with two kinds of problems. First, one has to deal with the basic,
conceptual questions arising when the underlying meson--baryon
field--theory is truncated to a limited number of particles (in
practice one resorts to a truncation to a finite number of
pions, since nucleon--antinucleon states are usually neglected).
Second, one is left with the non--trivial
question of how to get connected--kernel equations coupling all
the relevant transition amplitudes. For the $\pi$NN--NN case, this
problem has been solved by several authors following different
routes (for a thorough discussion
of $\pi$NN systems and the relevant literature see Ref. \cite{gm90}).
Whatever the approach may be, in any case, at the end
one is left with equations having more or less the same structure,
providing a connected--kernel scheme coupling scattering,
production and absorption amplitudes in a unified way. The situation
is far less satisfactory for the proper $\pi$--multinucleon problem.
To the best of our knowledge, for the $\pi$NNN--NNN case the only attempt
to a unitary, connected--kernel theory has been given by Avishai and
Mizutani \cite{am83} several years ago. These authors resorted to the
AGS approach in the quasi--particle approximation to get
connected--kernel equations in the pure four--body ($\pi$NNN) sector.
The disconnected contributions due to the coupling of the four--body
space to the three--nucleon one were then handled through a
two--potential formulation. This procedure,
although solving in principle the scattering problem, does not give a
closed set of coupled equations, and is too involved
to be considered for practical applications.
\par
      In this paper, we consider the
$\pi$--multinucleon problem with a completely new approach.
The key ingredient of the present formalism is
a proper generalization of the AGS matrix method to situations
in which elementary production or absorption $\pi$NN vertices
connect spaces with a different number of particles (the N--body space
made by the $\pi$ and $N$-1 nucleons, and the ($N$-1)--nucleon space).
The result is that the subsystem dynamics are naturally embedded
into the full pion--multinucleon problem, even when production/absorption
processes occur in the various subsystems.
This paper illustrates in full details the approach we have developed
in order to treat the pion--multinucleon problem.
It represents a complete article
discussing extensively the theory we have already outlined in a
recent Letter \cite{cc94} for the pion--trinucleon problem.
With respect to Ref. \cite{cc94}, however, this paper discusses the equations
in the more general framework of the pion--multinucleon problem, rather
than being limited to the specific pion--trinucleon system.

For a clear understanding of the theory, we consider indispensable
to revisit first the well--known results of the standard $N$--body theory.
Our revisitation has the advantage to maintain intelligible at every step
the connection between the physical amplitudes and the GS (chain--labelled)
$N$--body operators. In fact, our revisitation starts from one single
amplitude, which describes the fully unclusterized $N\leftarrow N$ process,
and bootstraps from there the whole $N$--body connected--kernel--equation
theory. It seems to us that this connection is more hardly achieved in
previous formulations.
Moreover, and what is most important, the simplicity of this formulation
allows to repeat in the second part of the paper a rather similar procedure
for the more complicated pion+($N$-1)--nucleon problem, where
pion absorption/production processes have to be taken into account.

In Sect. II, we consider the standard
few--body problem. Our procedure can be summarized as
follows. First we start from the $N\leftarrow N$ transition
amplitude describing $N$ incoming and outgoing particles, and
introduce the AGS transition operators referring to
$(N-1)$--cluster$\leftarrow (N-1)$--cluster transitions {\em via} a
suitable ansatz. Dynamical equations can be easily derived for these
operators starting from the Lippmann--Schwinger equation for the
$N\leftarrow N$ amplitude. Then one introduces operators labelled
by chains of partitions starting at the $(N-2)$--cluster level by
iterating the above procedure through the AGS matrix method. The
new (chain--labelled) operators can be related to the physical
transition amplitudes with $N-2$ clusters in the asymptotic states.
This procedure can be repeated, until $2$--cluster$\leftarrow
2$--cluster transitions are considered. These are associated to
operators labelled by complete chains, satisfying the well--known
AGS $N$--body equations, having a connected Yakubovski\u\i\ kernel.

In Sect. III, we start from the non--connected Afnan--Blankleider (AB)
equations coupling all the fully unclusterized
transition amplitudes one can have
in correspondence to one pion and $N$-1 nucleons \cite{ab80,cc93}.
These equations can be derived starting from meson--nucleon
field theory by means of Taylor' s diagrammatic method \cite{ta63,br69},
once diagrams with at most one pion are retained, and explicit multipion
intermediate states are disregarded.
Then, we introduce new chain--of--partition labelled operators
by exploiting the same ansatz technique already tested in the
standard $N$--body theory, and, from the AB equations,
we derive new dynamical equations for these operators.
It has to be observed that the extension of the AGS approach to
processes allowing
for production/absorption is in itself not trivial. This is most clearly
seen in the chain--labelling of the operators, which has to be modified
so as to deal simultaneously with spaces having a different number of
particles. (For example, in the $\pi$NNN--NNN case, one gets 24 coupled
equations, in place of the 18 equations of the standard Yakubovski\u\i--type
formalism.) For the general pion--multinucleon problem, our findings are not
exhaustive, in that the formalism does not complete the program for obtaining
connected--kernel equations, much in the same way the Yakubovski\u\i\
4--body equations are in fact connected only for $N$=4, while for $N >4$
they can be used as a starting point to move one step forward to the next
stage of the theory, which is connected for $N$=5, and so on. We conjecture
that the same hierarchic scheme could be attempted in order to achieve a
fully connected--kernel formulation of the general pion--multinucleon
problem, however this is beyond the limits of the present paper.
Finally, in the second part of Sect. III, we analyze in detail the
kernel of the equation we have obtained, and show for the first time that
such a kernel is indeed connected for the pion--trinucleon system.
As we will show, the connectedness properties of the coupled pion--trinucleon
equations are rather different from the standard Yakubovski\u\i\ four--body
equation. Whereas the standard four--body equations are connected after
two iterations, the presence of the meson--baryon vertex modifies
substantially the connectedness properties of the kernel, in that
it is achieved at different stages for different sectors
of the kernel: the $NNN\leftarrow NNN$ sector turns out to be connected
after one iteration, the $NNN\leftarrow NNN\pi$ after two, while
the remaining $NNN\pi\leftarrow NNN$ and $NNN\pi\leftarrow NNN\pi$
sectors require three iterations.

\section {$N$--body theory revisited}

\subsection{The fully unclusterized ($N\leftarrow N$) amplitude}
\label{subsec:uncluster}

Let's consider a system of $N$ particles interacting through
two--body potentials. If we consider the extreme scattering
situation of $N$ clusters going to $N$ clusters, where obviously each
cluster has to be made by one single particle,
the amplitude $T$ ($T$ $\equiv$ $T^{(N)}$ for brevity) for the process
$N$ $\leftarrow$ $N$ satisfies a $N$--body Lippmann--Schwinger
equation
\begin{equation}
T=V+VG_0T,
\label{LS}
\end{equation}
where $G_0$ is the $N$--body free resolvent operator, and the potential $V$
is defined by the sum of the two--body potentials pertaining to each pair $a$
\begin{equation}
V=\sum v_a.
\label{pairpot}
\end{equation}
If $N=2$, there is only one pair, and the solution of the
Lippmann--Schwinger equation uniquely determines the physical amplitude.

If N $> 2$, the solution of the Lippmann--Schwinger equation is not
uniquely determined, and therefore this equation
is not sufficient to find the
physical amplitude.
On the contrary, we can determine unambiguously the amplitude $t_a$
{\it internal} to a certain partition $a$ in $N-1$ clusters
\begin{equation}
t_a=v_a+v_aG_0t_a.
\label{2bodyLS}
\end{equation}
This is the two--body amplitude for the pair $a$, embedded in a
bath of $N-2$ spectators.

To determine the physical amplitude it is fundamental to rewrite
$T$ it in terms of new unknowns $U_{ab}$ using the following ansatz:
\begin{equation}
T=\sum_a t_a +\sum_{ab} t_aG_0U_{ab}G_0t_b.
\label{fundamental}
\end{equation}
The sums in Eq.~(\ref{fundamental}) range over all the possible
$N(N-1)/2$ pairs one can consider out of N particles. Therefore we
have introduced a new set of $N(N-1)/2 \times N(N-1)/2$ quantities $U_{ab}$.
To get the new equation for the unknown $U_{ab}$ we substitute
Eq.~(\ref{fundamental}) in Eq.~(\ref{LS}) and consider
Eq.~(\ref{2bodyLS}), thereby obtaining
\begin{equation}
\sum_{ab}t_aG_0U_{ab}G_0t_b=\sum_{ab}v_aG_0t_b\bar\delta_{ab}
+\sum_{acb}v_aG_0t_cG_0U_{cb}G_0t_b,
\label{perAGS-1}
\end{equation}
with $\bar\delta_{ab}$ defined as $1-\delta_{ab}$.
This equation will hold {\it a fortiori} if the unknowns $U_{ab}$
satisfy a similar term--by--term expression
\begin{equation}
t_aG_0U_{ab}G_0t_b=v_aG_0t_b\bar\delta_{ab}
+\sum_{c}v_aG_0t_cG_0U_{cb}G_0t_b.
\label{perAGS-2}
\end{equation}
Simple algebraic manipulations allow one to rewrite this expression
in the very famous form of Alt--Grassberger--Sandhas (AGS)
\cite{ags,gs67}
\begin{equation}
U_{ab}=G_0^{-1}\bar\delta_{ab}+\sum_c\bar\delta_{ac}t_cG_0U_{cb}.
\label{AGS}
\end{equation}
If N $=3$, the solution of the AGS equations is uniquely determined,
and hence the knowledge of the quantities $U_{ab}$ is sufficient to
determine the physical amplitude $T$ through Eq.~(\ref{fundamental}).

If N $> 3$ the program has not been concluded yet, since the
AGS equations are not sufficient to determine unambiguously
the quantities $U_{ab}$. However, we can follow Sandhas'
brilliant reinterpretation of the AGS equation and consider
it as a {\it super} Lippmann--Schwinger equation for matrices
defined in the space of pairs
\begin{equation}
 T^{(N-1)} ={V}^{(N-1)} +{V}^{(N-1)} {G_0}^{(N-1)} {T}^{(N-1)} ,
\label{SuperLS}
\end{equation}
with the definitions
\begin{eqnarray}
{T^{(N-1)} }_{ab}&\equiv& U_{ab} \label{Sandhasdef}\\
{V^{(N-1)} }_{ab}&\equiv& G_0^{-1}\bar\delta_{ab}\label{Sandhaspot}\\
{{G_0}^{(N-1)} }_{ab}&\equiv& G_0t_aG_0\delta_{ab}.\label{Sandhasprop}
\end{eqnarray}
Since we formally returned to a Lippmann--Schwinger equation
(but, as will be explained below, this {\it new} LS equation
gives the physical amplitudes for $(N-1)$ $\leftarrow$ $(N-1)$
clusters), we can repeat the same procedure outlined above
for the $T$ amplitude.

First, we introduce a new ``potential"
internal to a given ($N-2$)--cluster partition $a'$,
namely a matrix operator $v^{(N-1)}_{a'}$ $\equiv$
$G_0^{-1}\bar\delta_{ab}$, whose elements, labelled by
($N-1$)--cluster--partitions, are defined
only when the partitions $a$ and $b$ can be obtained from
$a'$ by breaking one of its clusters
($a\subset a'$ and $b\subset a'$). Then
the sum rule
\begin{equation}
{V^{(N-1)} } = \sum_{a'} v^{(N-1)}_{a'}
\label{triplepot}
\end{equation}
is straightforwardly proved
once it has been observed that two different pairs unambiguously
identify one single partition $a'$ in $N-2$ clusters \cite{ya67}.

Second, we consider the amplitude for an $N-1$ to $N-1$
process internal to the partition $a'$
in $N-2$ clusters ($a'$ can be either a 3-body cluster
embedded in a bath of $N-3$ spectators, or a couple of two distinct
pairs embedded in a bath of $N-4$ spectators), which is fully
determined by the equation
\begin{equation}
t^{(N-1)}_{a'}=v^{(N-1)}_{a'}+
v^{(N-1)}_{a'}G_0^{(N-1)}t^{(N-1)}_{a'}.
\label{N-2superLS}
\end{equation}
The notion of partition in $N-1$ clusters {\it internal}
to a given partition in $N-2$ clusters naturally brings us to the
introduction of chain of partitions. That is, out of all the
$N(N-1)/2$ partitions in $N-1$ clusters, only those coming
from a sequential break--up of one particular partition
${a'}$ of the
$N(N-1)(N-2)(3N-5)/24$ partitions in $N-2$ clusters are significant in
Eq.~(\ref{N-2superLS}). More generally, one can introduce
chains of partitions $\{a_{N-1}, a_{N-2},\ldots,a_{i+1},
a_{i}\} \equiv \alpha_i$, {\it i. e.} sets of partitions into $N-1,
N-2,\ldots,i+1, i$ clusters, where the partition into $j$ clusters
$a_j$ is obtained by the coalescence of two clusters of $a_{j+1}$
($a_{N-1}\subset a_{N-2}\subset \ldots \subset a_{i+1}\subset a_{i}$).

Finally, to get to the next equation (which is exhaustive if N=4)
we again rewrite the matrix
$T^{(N-1)}$ in terms of a set of new unknown matrices
$U^{(N-1)}_{a'b'}$ using the ansatz:
\begin{equation}
T^{(N-1)} =\sum_{a'} t^{(N-1)}_{a'} +\sum_{a'b'} t^{(N-1)}_{a'}
G_0^{(N-1)} U^{(N-1)}_{a'b'} G_0^{(N-1)} t^{(N-1)}_{b'},
\label{fundamental-1}
\end{equation}
where the sums range over all the possible partitions in $N-2$
clusters of an $N$--body system. The number of such unknown
matrices is the square of $N(N-1)(N-2)(3N-5)/24$.

Thus, we can view the set of new unknown
matrices  $U^{(N-1)}_{a'b'}$ as one {\it single}
matrix defined in the chain space spanned by elements
uniquely identified by one partition in $N-2$ clusters
and one partition in $N-1$ clusters {\it internal} to the
given partition in $N-2$ clusters.

Eqs.~(\ref{SuperLS}), (\ref{triplepot}), and (\ref{N-2superLS}) are
the counterparts of the perfectly similar Eqs.~(\ref{LS}),
(\ref{pairpot}), and (\ref{2bodyLS}), the difference being that
Eq.~(\ref{LS}) describes the scalar amplitude for the $N\leftarrow N$
process, while Eq.~(\ref{SuperLS}) describes the set of amplitudes for
all the $N-1\leftarrow N-1$ processes. Expression (\ref{fundamental})
was the fundamental ansatz one needed to obtain Eq.~(\ref{SuperLS})
starting from Eqs.~(\ref{LS}), (\ref{pairpot}), and (\ref{2bodyLS}).
Similarly, with the expression (\ref{fundamental-1}) it is now possible
to reach to the next stage of the hierarchy and derive the
4--body Yakubovski\u\i\ equations in the form of Grassberger--Sandhas.
The procedure to follow is the trivial replication of the one outlined
previously for the AGS equation. Instead of doing this, we will rather
exploit the recursive structure of this $N$--body theory and obtain the
general result which holds at {\it any} level of the hierarchy.

Suppose we have iterated ($n$-1) times the procedure outlined above,
thereby obtaining a ($n$-1)--{\it times super} LS
matrix equation
\begin{equation}
T^{(N-n+1)} =V^{(N-n+1)} +V^{(N-n+1)} G_0^{(N-n+1)} T^{(N-n+1)} ,
\label{LS-(N-n+1)}
\end{equation}
where the matrix elements of $T^{(N-n+1)}$ are labelled
by chains of partitions $\alpha_{N-n+1}$ and $\beta_{N-n+1}$
and refer to physical transition amplitudes between $(N-n+1)$ clusters
(see below).
Transition operators, potentials and Green's functions are
defined (for $n\geq 2$) as
\begin{eqnarray}
{T^{(N-n+1)}}_{ab}&\equiv&U^{(N-n+2)}_{ab} \label{tgs}\\
{V^{(N-n+1)}}_{ab} &\equiv&{G_0^{(N-n+2)}}^{-1}\bar\delta_{ab}
\label{vgs}\\
{G_0^{(N-n+1)}}_{ab} &\equiv&{G_0^{(N-n+2)}}
t^{(N-n+2)}_{a} {G_0^{(N-n+2)}}\delta_{ab} \label{pgs}
\end{eqnarray}
where we have denoted by $a,b$ two possible partitions
into $(N-n+1)$ clusters ($a\equiv a_{N-n+1}$ for brevity).

The quantities denoted with the superscript ${(N-n+1)}$ are matrices
labelled by the chain index $\alpha_{N-n+1}$ and the above equations
express them in terms of the ${(N-n+1)}$--cluster
partition index $a_{N-n+1}$ and operators with the superscript
${(N-n+2)}$. Obviously, these operators are labelled by the the internal
subchain index $\alpha_{N-n+2}$, but this is omitted for brevity.

If $n=N-1$, Eq.~(\ref{LS-(N-n+1)}) is connected and unambiguously defines
the unknown $T^{(N-n+1)}$. Since $T^{(N-n+1)}$ is the auxiliary
quantity for $T^{(N-n+2)}$, this also is defined, and hence
$T^{(N-n+3)}$ $\dots$ up to $T\equiv T^{(N)}$. Thus, our
initial problem is completely solved.

If $n< N-1$, first we recall the $(N-n)$--cluster partition sum
rule \cite{sa80,cv79} (for brevity $a'\equiv a_{N-n}$)
\begin{equation}
{V^{(N-n+1)}} = \sum_{a'} v^{(N-n+1)}_{a'} .
\label{n-pot}
\end{equation}
Second we introduce the amplitude internal
to the partition $a_{N-n}$, which is unambiguously defined
through the equation
\begin{equation}
t^{(N-n+1)}_{a'} =
v^{(N-n+1)}_{a'} + v^{(N-n+1)}_{a'} G_0^{(N-n+1)}
t^{(N-n+1)}_{a'}.
\label{int-LS-(N-n+1)}
\end{equation}

Third, we introduce the {\it nth--step} auxiliary unknowns
\begin{equation}
{T^{(N-n+1)} }=\sum_{a'} t^{(N-n+1)}_{a'} +\sum_{a'b'}
t^{(N-n+1)}_{a'} {G_0^{(N-n+1)}}
U^{(N-n+1)}_{a'b'}{G_0^{(N-n+1)}} t^{(N-n+1)}_{b'}.
\label{fundamental-n}
\end{equation}

Substitution of Eq~.(\ref{fundamental-n}) into Eq.~(\ref{LS-(N-n+1)}),
with the use of Eq.~(\ref{int-LS-(N-n+1)}), and with the identification
{\it term by term} of the resulting equation, yields
directly the famous Yakubovski\u\i--Grassberger--Sandhas (YGS)
$N$--body equations \cite{gs67,agsp}
\begin{equation}
U^{(N-n+1)}_{a'b'} =\bar\delta_{a'b'}{G_0^{(N-n+1)}}^{-1}+
\sum_{c'}\bar\delta_{a'c'}t^{(N-n+1)}_{c'}
{G_0^{(N-n+1)}} U^{(N-n+1)}_{c'b'}. \label{ags-(N-n+1)}
\end{equation}
If $n=N-1$ this equation is fully connected and uniquely defines the
auxiliary unknowns. Otherwise it can be recast into a new super
Lippann--Schwinger equation and used to move another step forward.

\subsection{The clusterized transition amplitudes}
\label{subsec:cluster}
      To relate the operators $U_{a'b'}^{(N-n+1)}$ to the physical
transition amplitudes, we consider first Eq.~(\ref{fundamental}),
which expresses the $N\leftarrow N$ amplitude $T$ in terms
of the operators $U_{ab} \equiv {T^{(N-1)} }_{ab}$. The $N - 1 \leftarrow
N - 1$ transition amplitudes can be obtained from the plane--wave
representation of Eq.~(\ref{fundamental})
by recalling that, near a bound--state pole at energy $E_a$, the
two--body $t$--matrix $t_a$ can be written in the separable form
\begin{equation}
t_a \simeq {v_a |\Phi_{a}><\Phi_{a}| v_a \over z - E_{a}} ,
\label{eq:sep1}
\end{equation}
with $|\Phi_a>$ the channel state describing the bound pair $a$
plus $N - 2$ non--interacting particles. It obeys the homogeneous
equation
\begin{equation}
G_0(E_a) v_a |\Phi_a> = |\Phi_a>.
\label{eq:homo1}
\end{equation}
Extracting from Eq. (2.4) the left and right residues at the
appropriate poles one then has
\begin{equation}
{\cal T}_{ab} \equiv <\Phi_a|v_a G_0 U_{ab} G_0 v_b |\Phi_b> =
<\Phi_a|U_{ab}|\Phi_b>
\label{eq:amp1}
\end{equation}
as the physical transition amplitudes for the $N - 1 \leftarrow N - 1$
transitions.
\par
      The above considerations can be extended to $N - 2 \leftarrow
N - 2$ transitions and to the operators $U_{aa'\ bb'}$, labelled
by (partial) chains of partitions by considering
Eq.~(\ref{fundamental-1}). An
asymptotic configuration of energy $E_{a'}$, describing either a
three--body bound state plus $N-3$ free particles, or two two--body bound
states plus $N - 4$ non--interacting objects, is now associated to the
channel state $|\Phi_{a'}>$, whose Faddeev components
\begin{equation}
|\Phi_{a;\ a'}> \equiv G_0 v_a |\Phi_{a'}>
\label{eq:comp1}
\end{equation}
satisfy the homogeneous Faddeev equations
\begin{equation}
|\Phi_{a;\ a'}> = G_0 t_a \sum_{c (\subset a')} \bar \delta_{ac}
|\Phi_{c;\ a'}> .
\label{eq:homofa1}
\end{equation}
Because of the very definition of $v_{a'}^{(N-1)}$ and $G_0^{(N-1)}$
(see Eqs. (\ref{Sandhaspot}), (\ref{Sandhasprop}) and (\ref{triplepot}))
the Eqs.~(\ref{eq:homofa1}) can be written in a form analogous to Eq.
(\ref{eq:homo1}), namely
\begin{equation}
G_0^{(N-1)} (E_{a'}) v_{a'}^{(N-1)}|\tilde \Phi_{a'}> = |\tilde \Phi_{a'}>
.\label{eq:homo2}
\end{equation}
Here, $|\tilde \Phi_{a'}>$ represents a column--vector with components
$|\Phi_{a;\ a'}>$. Much as in the usual two--body problem, one can
regard Eq.~(\ref{eq:homo2}) as an effective "channel" Schr\"odinger
equation
\begin{equation}
(G_{0}^{(N-1) -1}(E_{a'}) - v_{a'}^{(N-1)})|\tilde \Phi_{a'}> = 0 ,
\label{eq:sch}
\end{equation}
whose associated T--matrix $t_{a'}^{(N-1)}$ has a pole at the energy
$E_{a'}$. In other words one can write for $z \simeq E_{a'}$
\begin{equation}
t_{a'}^{(N-1)} \simeq {v_{a'}^{(N-1)}|\tilde \Phi_{a'}>
<\tilde \Phi_{a'}| v_{a'}^{(N-1)} \over z - E_{a'}} \ ,
\label{eq:sep2}
\end{equation}
with $<\tilde \Phi_{a'}|$ a row--vector satisfying
\begin{equation}
<\tilde \Phi_{a'}| = <\tilde \Phi_{a'}| v_{a'}^{(N-1)} G_0^{(N-1)}
(E_{a'}) \ .
\end{equation}
Using Eq.~(\ref{eq:sep2}) in Eq.~(\ref{fundamental-1}) to extract
the left and right residues one has that the
physical $N -2 \leftarrow N -2$ transition amplitudes are given by
\begin{eqnarray}
{\cal T}_{a'b'}^{(N-2)} & = <\tilde \Phi_{a'}| v_{a'}^{(N-1)} G_0^{(N-1)}
U_{a'b'}^{(N-1)} G_0^{(N-1)} v_{b'}^{(N-1)}
|\tilde \Phi_{b'}> \nonumber \\
& = <\tilde \Phi_{a'}| U_{a'b'}^{(N-1)} |\tilde \Phi_{b'}> .
\label{eq:amp2}
\end{eqnarray}
This result can be written in more explicit terms as follows:
\begin{equation}
{\cal T}_{a'b'}^{(N-2)} = \sum_{ab} <\Phi_{a;\ a'}|U_{aa'\ bb'}
|\Phi_{b;\ b'}> .
\label{eq:amp2p}
\end{equation}
\par
      The generalization to the $n$--th step of the inductive procedure
is now straightforward. The homogeneous equation (\ref{eq:homo2})
is now replaced by
\begin{equation}
G^{(N - n + 1)}_0 v^{(N - n + 1)}_{a'} |\tilde \Phi_{a'}> =
|\tilde \Phi_{a'}>
\label{eq:homon}
\end{equation}
where $2\leq n \leq N - 2$, and $|\tilde \Phi_{a'}>$ is a column vector
having the Faddeev--Yakubovski\u\i\ components \cite{cv83f}
\begin{equation}
|\Phi_{\alpha_{N-n+1};\ a'}> \equiv
G_0v_{a_{N-1}}G_{a_{N-1}}\ldots G_{a_{N-n+1}} v^{a_{N-n+1}}_{a'}|\Phi_{a'}>
\label{eq:fycompo}
\end{equation}
of the channel state $|\Phi_{a'}>$ as its elements. As well known,
the components $|\Phi_{\alpha_{N-n+1};\ a'}>$ sum up to give
the full channel state $|\Phi_{a'}>$ \cite{cv83f,cv79b}. Eq.~(\ref{eq:homon})
is just the matrix version of the homogeneous Faddeev--Yakubovski\u\i\
equations for these components. Here, $v_{a_i}^{a_j}$ represents the
interaction internal to partition $a_i$ and external to $a_j$
$(a_j \subset a_i)$, and $G_{a_i}$ is the resolvent associated to
$a_i$.
\par
      Near a bound--state pole $E_{a'}$ the matrix operator
$t_{a'}^{(N-n+1)}$ can be then approximated by its dominant term
\begin{equation}
t^{(N-n+1)}_{a'} \simeq {v^{(N-n+1)}_{a'} |\tilde \Phi_{a'}>
<\tilde \Phi_{a'}| v^{(N-n+1)}_{a'} \over z - E_{a'}}
\label{eq:sepn}
\end{equation}
which is the counterpart of Eq.~(\ref{eq:sep2}) for the $n$th--step
case. Extraction of the appropriate residues from (\ref{fundamental-n})
gives
\begin{eqnarray}
{\cal T}^{(N-n)}_{a'b'} & = <\tilde \Phi_{a'}| v^{(N-n+1)}_{a'}
G^{(N-n+1)}_0 U^{(N-n+1)}_{a'b'} G^{(N-n+1)}_0
v^{(N-n+1)}_{b'} |\tilde \Phi_{b'}> \nonumber \\
& = <\tilde \Phi_{a'}| U^{(N-n+1)}_{a'b'} |\tilde \Phi_{b'}>
\label{eq:ampn}
\end{eqnarray}
as physical transition amplitudes for $(N - n)\leftarrow(N - n)$
transitions. We remind that the matrices in Eq.~(\ref{eq:ampn})
have to be regarded as labelled by {\it chains} of partitions
starting at the $(N - n)$th--cluster level from either partition
$a' \equiv a_{N -n}$, or partition $b' \equiv b_{N -n}$, so that
Eq. (\ref{eq:ampn}) explicitly reads
\begin{equation}
{\cal T}^{(N-n)}_{a'b'} = \sum_{\stackrel{a_{N-1}\ldots a_{N-n+1}}
{b_{N-1}\ldots b_{N-n+1}}} <\Phi_{a_{N-1}\ldots a_{N-n+1};\ a'}|
U^{(N-n+1)}_{a_{N-1}\ldots a'\ b_{N-1}\ldots b'}
|\Phi_{b_{N-1}\ldots b_{N-n+1};\ b'}> .
\label{eq:ampnp}
\end{equation}
In particular, for $n = N - 2$ this result provides the $2\leftarrow 2$
transition amplitudes in terms of the chain--labelled GS operators
$U_{a_{N-1}\ldots a_2\ b_{N-1}\ldots b_2}$ satisfying connected--kernel
YGS equations \cite{note2}.

\section {The $\pi$--multinucleon case}

  In this Section we consider how the original GS approach to the
$N$--body problem has to be modified, when dealing with a system
consisting of a pion and $N$-1 nucleons. This generalization if far
from trivial, because the absorption process requires the simultaneous
and consistent treatment of the $N$--body and ($N$-1)--nucleon continua.
The fact that the number of particles is not conserved pushes the theory
well beyond the standard $N$--body scheme. These
difficulties are clearly perceivable in the paper by Avishai and
Mizutani \cite{am83}, which, to our knowledge at least, represents the only
complete attempt to generalize the GS approach to
$\pi$--multinucleon dynamics. There, the original GS quasi--particle
formulation is employed, so as to recast the $\pi$NNN--NNN problem into
an effective multichannel problem, coupling isobars to the
corresponding three--nucleon space. Connectedness is achieved in the
four--body sector {\it via} the GS procedure, the complexities due
to pion production/absorption being successively taken into account
through a two--potential formulation. This route makes the whole
formalism much more involved than the original GS approach.
Here, we show that the basic GS strategy can be followed much more
faithfully, so as to provide the natural extensions of the "potential"
sum rule (\ref{triplepot}), and a more correct embedding of the
subsystem dynamics into the $\pi$--multinucleon theory.
This leads to consistent generalizations
of the GS matrix--operators (\ref{tgs}), (\ref{vgs}) and (\ref{pgs}),
and to a closed set of equations coupling all the
relevant scattering and production/absorption operators.

\subsection{General scheme}

      As in the previous Section, we start from fully unclusterized
amplitudes. However, in addition to the $N\leftarrow N$ amplitude
$T_{(1|1)}$, with $N$-1 nucleons and one physical pion in the asymptotic
region, here we have also to consider the amplitude $T_{(0|0)}$ for
the ($N$-1)--nucleon scattering processes, as well as the the fully
unclusterized absorption and production amplitudes, with no pion
in the final or initial channel, respectively. We denote the former
as $T_{(0|1)}$ and the latter as $T_{(1|0)}$. These physical fully
unclusterized amplitudes are connected to the solutions of the
Afnan--Blankleider equations through the relations:

\begin{mathletters}
\begin{eqnarray}
T_{(1|1)}&=&\sum_a t_a +\sum_{ab} t_aG_0U_{ab}G_0t_b,
\label{T}\\
T_{(0|1)}&=&\sum_{a} U^\dagger_{a}G_0t_a,\\
T_{(1|0)}&=&\sum_{a} t_aG_0U_{a},\\
T_{(0|0)}&=& U.
\end{eqnarray}
\end{mathletters}

The auxiliary amplitudes $U_{ab}, U^\dagger_a, U_a, U$
satisfy the Afnan--Blankleider coupled equations \cite{am83,ab80,cc93}
\begin{mathletters}
\begin{eqnarray}
U_{ab}&=&G_0^{-1}\bar\delta_{ab}+\sum_c\bar\delta_{ac}t_cG_0U_{cb}
+F_a g_0 U^\dagger_b, \label{ABa}\\
U^\dagger_a&=&F^\dagger_a+{\cal V}g_0U^\dagger_a+\sum_c F^\dagger_c
G_0t_cG_0U_{ca},\label{ABb}\\
U_a&=&F_a+\sum_c\bar\delta_{ac}t_cG_0U_c+F_ag_0U,\label{ABc}\\
U&=&{\cal V}+{\cal V}g_0U+\sum_c F^\dagger_c G_0t_cG_0U_c .\label{ABd}
\end{eqnarray}
\end{mathletters}
These equations couple together the fully unclusterized
amplitude for the $(N-1)\leftarrow (N-1)$ process {\it without pions}
with all the $(N-1)$--cluster amplitudes $U_{ab}$ one can have
{\it with the pion}. In other words, the set of equations for
$U_{ab}, U^\dagger_a, U_a, U$ couple together
the amplitudes for {\it all} ({\it i.e.}, with or without the pion!)
the partitions into $N$-1 clusters one can have out of a system made
by one pion and $N$-1 nucleons. The number of such partitions is
${N(N-1)\over 2}+1$.

      In Eqs.~(\ref{ABa})--(\ref{ABd}), in addition to the $N$--body
free propagator $G_0$, the $(N-1)$--nucleon propagator $g_0$ appears.
It takes into account the $(N-1)$--particle free motion in the pure
multinucleon sector. As for the vertex operators $F_a$ and $F^{\dagger}_a$,
labelled by the ($N$-1)--cluster index $a$, they are related to the basic
$\pi{\rm NN}$ vertices $f(i)$ and $f^{\dagger}(i)$, referring to the $i$--th
nucleon, by
\begin{eqnarray}
F_a&=&\sum_{i=1}^{N-1}\bar\delta_{ia} f(i),\\
F^\dagger_a&=&\sum_{i=1}^{N-1}\bar\delta_{ia} f^\dagger(i),
\end{eqnarray}
for production and absorption, respectively. Because of their importance,
these positions deserve further comments. The index $a$, being the
$(N-1)$--cluster label of the $\pi$--multinucleon ($N$--body) sector, denotes
also all the pairs one can make out of $N$ bodies. The index $i$ labels each
one of the $N-1$ nucleons; however, it can also be used to denote the
specific pair ($\pi {\rm N}_i$). Thus, if $a=({\rm NN})$, $F_a$ is given by
the sum of {\it all} the $\pi {\rm NN}$ vertices; if $a=(\pi {\rm N}_i)\equiv
i$, the vertex $\pi {\rm N}_i {\rm N}_i$ has to be excluded in
the sum. We recall that $f(i)$ represents a {\it dressed} form factor.
In time--ordered perturbation theory, it can be written in terms of the
bare vertex $f_0(i)$ as $f(i) = (1+t_iG_0)f_0(i)$ \cite{ab80,cc93}.
Similar considerations apply to the absorption vertex $f^{\dagger}(i)$.

When dealing with a theory involving production/absorption processes,
care must be exercised in the treatment of the two--body interactions.
We recall that the Eqs.~(\ref{ABa})--(\ref{ABd}) emerge, once the
$\pi {\rm N}$ and NN interactions are approximated with static
potentials in the pure $N$--body sector. Since pion emission and absorption
are explicitly considered in the $N \leftarrow N$ transition amplitude
$T_{(1|1)}$,
the polar $P_{11}$ term has to be excluded in the $\pi$--nucleon
t--matrices $t_i$. The NN transition operators, on the other hand,
must refer to the full nucleon--nucleon interaction.
\par
      In principle, the potential ${\cal V}$ in Eqs.~(\ref{ABb}) and
(\ref{ABd}) consists of all the two--body one--pion--exchange interactions,
which are included in the theory through explicit pion emission,
propagation and absorption, plus two--body heavy--boson--exchange
interactions and multinucleon forces. These contributions take
into account in an effective way the intermediate multipion states,
which are not properly included in the Afnan--Blankleider approach,
(we recall that in this theory only states with at most {\it one} pion
are explicitly considered). Since here we are mainly interested in
the derivation of connected--kernel equations of the GS type, we do not
address the question of an explicit treatment of multibody forces.
Formally, in any case, these contributions can be always distributed
as sums
over two--body terms, in correspondence to the possible NN pairs.
\par
      The approach to coupled $\pi$--multinucleon systems we are
considering here is not free from  fundamental conceptual problems.
These difficulties are directly related to the truncation of the
underlying field theory one has to perform to get coupled linear
equations referring to a finite number of degrees of freedom.
Because of this, it has been already pointed out that the
$\pi$NN coupling constant
in the nuclear environment turns out to be smaller than for an
isolated nucleon \cite{ss85,b92}.
Also the dressing of the free multi--nucleon propagator $g_0$
has problems. Indeed, graphs
where two (or more) pions dress the fermion lines at the same time
have to be excluded in $g_0$. The effect becomes more severe
by increasing the number of nucleons.
A related problem concerns the differences in dressing the nucleon
lines in the two propagators $G_0$ and $g_0$. Truncation at the level
of one--pion diagrams  leads to a differentiation of the nucleon masses
for the two sectors. Although some of these problems can be
technically recovered, or at least minimized,
these conceptual difficulties are presumably unavoidable
when the underling theory (which should in principle deal with an infinite
number of degrees of freedom) is forced to lay on the Procrustean bed
of non--relativistic few--body quantum mechanics.
This is the acknowledged difficulty of the approach, which is still
subject of thorough investigations even for the simpler $\pi$NN
case. We shall not dwell upon these problems any longer and refer the
interested reader to the relevant literature \cite{kb93,pa93,pa94,bk94}.

Here, we accept the approach as described
by Eqs.~(\ref{ABa})--(\ref{ABd}) as {\it starting point},
and address our attention to the insertion of such an approach
into a more general scheme required when dealing with more than two
nucleons.
Indeed, for two nucleons (N=3), the Afnan--Blankleider (AB) scheme is
fully connected and it is possible to evaluate without ambiguities all
the amplitudes into two and, through Eqs. (3.1), also into three
clusters.

If N$\geq$4 we proceed in close analogy to the standard $N$--body
case, namely we formally rewrite the AB equation
as a {\it super}--Lippmann--Schwinger equation for matrices defined
in {\it all} the partitions in $N-1$ clusters of the system.
\begin{equation}
{\sf T}^{(N-1)} ={\sf V}^{(N-1)} +
{\sf V}^{(N-1)} {\sf G}_{0}^{(N-1)} {\sf T}^{(N-1)} ,
\label{(N-1)-LS}
\end{equation}
with the definitions
\begin{eqnarray}
{{\sf T}}^{(N-1)}
\equiv
& \left|
\begin{array}{cc}
{T} ^{(N-1)}_{(a|b)} & {T} ^{(N-1)}_{(a|0)} \cr
{T} ^{(N-1)}_{(0|b)} & {T} ^{(N-1)}_{(0|0)} \cr
\end{array} \right|
&=
\left| \begin{array}{ccc}
U_{ab} &\ & U_a\\
U^\dagger_b &\ & U \\
\end{array} \right|, \label{newt}
\\
{{\sf V}}^{(N-1)}
\equiv & \left|
\begin{array}{cc}
{V} ^{(N-1)}_{(a|b)} & {V} ^{(N-1)}_{(a|0)} \cr
{V} ^{(N-1)}_{(0|b)} & {V} ^{(N-1)}_{(0|0)} \cr
\end{array} \right|
&=
\left| \begin{array}{ccc}
{G_0}^{-1}\bar\delta_{ab} &\ & F_a\\
F^\dagger_b &\ &{\cal V} \\
\end{array} \right|, \label{newv}
\\
{{\sf G}_{0}}^{(N-1)}
\equiv&
\left|
\begin{array}{cc}
{G_{0}} ^{(N-1)}_{(a|b)} & {G_{0}} ^{(N-1)}_{(a|0)} \cr
{G_{0}} ^{(N-1)}_{(0|b)} & {G_{0}} ^{(N-1)}_{(0|0)} \cr
\end{array} \right|
&=
\left| \begin{array}{ccc}
G_0t_aG_0\delta_{ab} &\ & 0 \\
0 &\ & g_0 \\
\end{array} \right|.
\label{newg-o}
\end{eqnarray}

It is worth to observe that, in the pure $N$--body sector, these
definitions are the same as in the standard GS formulation (compare
Eqs. (\ref{newt}), (\ref{newv}) and (\ref{newg-o}) with Eqs.
(\ref{Sandhasdef}), (\ref{Sandhaspot}) and (\ref{Sandhasprop}),
respectively)). However, due to the presence of the $\pi$NN vertex,
these matrix--operators acquire now extra elements corresponding to
the absorption/production processes, and to the pure nucleonic sector.

      As next step, we introduce in the system the partitions into
$N-2$ clusters. In the $N$--body sector, we again denote these partitions
by $a'$, while in the $(N-1)$--nucleon sector, we use the label $a_1$.
Clearly, $a_1$ denotes a pair of nucleons out of $N-1$ nucleons (without
the pion). Since the $N$--body sector can communicate with the
$(N-1)$--nucleon space, a further classification of the $(N-2)$--cluster
partitions $a'$ is now necessary, depending upon the way
the pion is associated to the composite clusters. Indeed, one can divide
these partitions into 4 classes. {Type--I} denotes a cluster ($\pi$NN)
embedded in a bath of $N-3$ purely spectator nucleons. {Type--II}
denotes a couple of pairs ($\pi$ N), (NN), embedded in a bath of $N-4$
spectator nucleons. {Type--III} represents a cluster (NNN)
embedded in a bath of $N-3$ pure spectators ($N-4$ nucleons and
the pion). Finally, {Type--IV} denotes a couple of pairs
(NN), (NN), embedded in a bath of $N-4$ pure spectators
($N-5$ nucleons and the pion). Obviously, type--IV partitions
occur only if $N\geq 5$.

{}From the standard $N$--body theory, it is obvious \cite{sa80,cv79}
that
\begin{mathletters}
\begin{eqnarray}
(V^{(N-1)})_{ab} &=&\sum_{a'} (v^{(N-1)}_{a'})_{ab}
\label{sumrule} \\
{\cal V} &=&\sum_{a_1} {\cal V}_{a_1}
\label{sumruleb} \end{eqnarray}
for the separated $N$--body and (N-1)--body sectors, respectively.

It is an important observation to realize that a similar
($N$-2)--cluster sum rule holds also for the corresponding vertices:
\begin{eqnarray}
F_{a}=\sum_{a'} (f_{a'})_{a} \\
F^{\dagger}_{a}=\sum_{a'} (f^{\dagger}_{a'})_{a} , \label{sumrulec}
\end{eqnarray}
\end{mathletters}
with the internal vertex operator defined as
\begin{equation}
(f_{a'})_{a} = \sum_{i=1}^{N-1}\bar\delta_{ia} f(i)\qquad {\rm with}
\qquad i, a \subset a' , \label{intvert}
\end{equation}
and similarly for $(f^\dagger_{a'})_{a}$. This is proved very easily once it
has been observed that two different pairs ($a,i$) unambiguously
identify one single partition $a'$ into $N-2$ clusters that contains
both of them.

We may also add that, if the pairs ($a,i$) have one
particle in common (including the pion) then $a'$ is of type--I,
while it is of type--II otherwise. Therefore there is an internal
vertex, and hence a coupling to the absorption channel, only for
partitions of type--I and --II. This has very important consequences
because only such two types of partitions unambiguously
identify one single nucleon--nucleon pair $a_1$ in the $(N-1)$--body
sector.

For each of the four types of $(N-2)$--cluster partitions
we have to introduce the corresponding sub--amplitudes.

{\bf Type--III and --IV)} Here, since the vertex operators are not effective,
we have standard $N$--body--like sub--amplitudes.
To fix the ideas, in the $\pi$--NNNN system we would have
sub--amplitudes such as (3N)+1N+$\pi$ and (2N)+(2N)+$\pi$, respectively.
Both III-- and IV--type sub--amplitudes are well--defined through
an AGS equation of the form (\ref{N-2superLS}), namely
\begin{eqnarray}
(u_{a'})_{ab}&=&G_0^{-1}\bar\delta_{ab}
+\sum_c\bar\delta_{ac}t_cG_0(u_{a'})_{cb}
\label{III-AB}\end{eqnarray}
with ${a,b,c\subset a'}$.

{\bf Type--I)} Because of the coupling with the $(N-1)$--nucleon space,
we do not have here a standard AGS equation. Rather, we have an
AB equation with corresponding sub--amplitudes
$(u_{a'})_{ab}, (u^\dagger_{a'})_{a}$ and $(u_{a'})_{a},
(u_{a'})_{}$. These equations couple together all the sub--amplitudes
for the $(N-1)$--cluster partitions {\it including} the absorption sector
when partitioned in $N-1$ single nucleons. We observe that, for a given
type--I partition $a'$, there is a unique nucleon--nucleon pair $a_1$
internal to the given  $a'$. Thus, we get the type--I sub--amplitudes
from the connected--kernel coupled equations \cite{ab80,cc93}
\begin{eqnarray}
(u_{a'})_{ab}&=&G_0^{-1}\bar\delta_{ab}
+\sum_c\bar\delta_{ac}t_cG_0(u_{a'})_{cb}
+(f_{a'})_a g_0 (u^\dagger_{a'})_{b}, \label{I-ABpp}
\\
(u^\dagger_{a'})_{a}&=&(f_{a'}^\dagger)_a+{\cal V}_{a_1} g_0
(u^\dagger_{a'})_{a}+\sum_c (f_{a'}^\dagger)_c G_0t_cG_0(u_{a'})_{ca},
\\
(u_{a'})_{a}&=&(f_{a'})_a+\sum_c\bar\delta_{ac}t_cG_0
(u_{a'})_{c}+(f_{a'})_ag_0(u_{a'})_{},
\\
(u_{a'})_{}&=&{\cal V}_{a_1}+{\cal V}_{a_1}g_0
(u_{a'})_{}+\sum_c(f^\dagger_{a'})_cG_0t_cG_0(u_{a'})_{c} ,
\label{I-AB}\end{eqnarray}
with ${a,b,c\subset a'}$, and {$a_1\leftarrow a'$}.

{\bf Type--II)} Like the I--type ones, these
sub--amplitudes couple the $N$--body sector with the $(N-1)$--nucleon one.
However, the equations for the II--type sub--amplitudes are
somehow different.
In fact, there is a delicate double--counting
problem connected with the introduction of such sub--amplitudes,
which requires a nontrivial modification with respect to the above
I--type equations. Let's observe the sum rule (\ref{sumruleb})
for the $(N-1)$--body sector.
This sum rule has been completely exhausted since it has been
already used for the AB equation for the I--type sub--amplitudes, and
hence, the equations for the II--type sub--amplitudes must take into
account this fact:
\begin{eqnarray}
(u_{a'})_{ab}&=&G_0^{-1}\bar\delta_{ab}
+\sum_c\bar\delta_{ac}t_cG_0(u_{a'})_{cb}
+(f_{a'})_a g_0 (u^\dagger_{a'})_{b}, \label{II-ABpp}
\\
(u^\dagger_{a'})_{a}&=&(f_{a'}^\dagger)_a+
\sum_c (f_{a'}^\dagger)_cG_0t_cG_0(u_{a'})_{ca},
\\
(u_{a'})_{a}&=&(f_{a'})_a+\sum_c\bar\delta_{ac}t_cG_0
(u_{a'})_{c}+(f_{a'})_ag_0(u_{a'})_{},
\\
(u_{a'})_{}&=&\sum_c(f^\dagger_{a'})_cG_0t_cG_0(u_{a'})_{c} .
\label{II-AB}\end{eqnarray}
Obviously, ${a,b,c\subset a'}$, and ${a_1\leftarrow a'}$.

For any type (I, II, III, and IV) of $(N-2)$--cluster partitions
of the $N$--body sector,
the above equations can be formally recast into
a {\it super}--Lippmann--Schwinger matrix equation,
namely
\begin{equation}
{\sf t}^{(N-1)}_{a'}
={\sf v}^{(N-1)}_{a'}
+{\sf v}^{(N-1)}_{a'}
{\sf G}_0^{(N-1)}
{\sf t}^{(N-1)}_{a'},
\label{(N-1)-subLS}\end{equation}
with ${\sf G}_0^{(N-1)}$ defined in Eq.~(\ref{newg-o}).

Since the new T--matrices ${\sf t}^{(N-1)}_{a'}$ are defined
in the space of all the $(N-1)$--cluster partitions,
their definitions depend upon whether ${a'}$ is of type I and II,
or III and IV. In fact, we have
\begin{eqnarray}
{\sf t}_{a'}^{(N-1)}
&\equiv \left|
\begin{array}{cc}
{t_{a'}} ^{(N-1)}_{(a|b)} & {t_{a'}} ^{(N-1)}_{(a|0)} \cr
 {t_{a'}} ^{(N-1)}_{(0|b)} &{t_{a'}}^{(N-1)}_{(0|0)}  \cr
\end{array} \right| = \left| \begin{array}{cc}
(u_{a'})_{ab} & (u_{a'})_{a}\\
({u^\dagger}_{a'})_{b}& u_{a'} \\
\end{array} \right|,
\end{eqnarray}
\begin{eqnarray}
{\sf t}_{a'}^{(N-1)}
\equiv \left|
\begin{array}{c}
{t_{a'}} ^{(N-1)}_{(a|b)} \cr
\end{array} \right| =
\left| \begin{array}{c}
{(u_{a'})}_{ab} \\
\end{array} \right|,
\end{eqnarray}
for partitions of type (I, II) or (III, IV) respectively.

For III-- and IV--type partitions the identification of
${\sf v}_{a'}$ corresponds to the standard
$N$--body choice in the $N$--body sector,
because there is no coupling with the $(N-1)$--nucleon sector. Hence one
has (see Eqs.~(\ref{Sandhaspot}) and (\ref{triplepot}))
\begin{eqnarray}
{\sf v}_{a'}^{(N-1)}
\equiv \left|
\begin{array}{c}
{v_{a'}} ^{(N-1)}_{(a|b)} \cr
\end{array} \right| =
\left| \begin{array}{c}
{G_0}^{-1}\bar\delta_{ab}\delta_{a,b\subset a'} \\
\end{array} \right| .
\end{eqnarray}

For I--type partitions, ${\sf v}_{a'}$
has 4 non null sectors
\begin{eqnarray}
{\sf v}_{a'}^{(N-1)}
\equiv \left|
\begin{array}{cc}
{v_{a'}} ^{(N-1)}_{(a|b)} & {v_{a'}} ^{(N-1)}_{(a|0)} \cr
 {v_{a'}} ^{(N-1)}_{(0|b)} & {v_{a'}}^{(N-1)}_{(0|0)} \cr
\end{array} \right| =
\left| \begin{array}{cc}
{G_0}^{-1}\bar\delta_{ab}\delta_{a,b\subset a'} &
(f_{a'})_{a}\\
({f^\dagger}_{a'})_{b}&{\cal V}_{a_1} \\
\end{array} \right| .
\end{eqnarray}

For II--type partitions, ${\sf v}_{a'}$ is defined
according to Eq.~(\ref{II-AB}):
\begin{eqnarray}
{\sf v}_{a'}^{(N-1)}
&\equiv \left|
\begin{array}{cc}
{v_{a'}} ^{(N-1)}_{(a|b)} & {v_{a'}} ^{(N-1)}_{(a|0)} \cr
 {v_{a'}} ^{(N-1)}_{(0|b)} &{v_{a'}}^{(N-1)}_{(0|0)}  \cr
\end{array} \right| = \left| \begin{array}{cc}
{G_0}^{-1}\bar\delta_{ab}\delta_{a,b\subset a'} &
(f_{a'})_{a}\\
({f^\dagger}_{a'})_{b}&0 \\
\end{array} \right|.
\end{eqnarray}

Thus, the sum rules (3.9) can be formally replaced by the single
{\it super} sum rule
\begin{equation}
{\sf V}^{(N-1)}=\sum_{a'}{\sf v}^{(N-1)}_{a'}.
\label{supersumrule}\end{equation}

Eqs.~(\ref{(N-1)-LS}), (\ref{(N-1)-subLS}), and (\ref{supersumrule})
are all the basic ingredients we need to derive a new equation which,
at least when $N=4$, yields a unique solution. The solution of such an
equation is then used to calculate the amplitudes ${\sf T}^{(N-1)}$.
On the contrary, a straightforward solution of the original AB equations
would lead to non unique results since for $N\geq$4 the AB kernel is not
connected. The path we have to follow closely mimics the $N$--body
procedure we have illustrated in the previous Section. The fundamental
ansatz
\begin{equation}
{{\sf T}^{(N-1)} }=\sum_{a'}
{\sf t}^{(N-1)}_{a'} +\sum_{a'b'}
{\sf t}^{(N-1)}_{a'}
{{\sf G}_0}^{(N-1)}
{\sf U}^{(N-1)}_{a'b'}
{{\sf G}_0}^{(N-1)}
{\sf t}^{(N-1)}_{b'},
\label{newfundamental}
\end{equation}
allows one to write the AB amplitudes in terms of new unknowns
${\sf U}^{(N-1)}_{a'b'}$. Again, we can regard the set of matrices
${\sf U}^{(N-1)}_{a'b'}$ as one single matrix defined in the chain
space labelled by one partition into $N-2$ clusters and one $(N-1)$--cluster
partition internal to the given $N-2$ clusterization. Contrary to
what happens in the standard $N$--body theory, however, here the absorption
channel has to be explicitly included. Thus, if one considers, for example,
the $\pi$NNN system there are 24 components rather than the
18 Yakubovski\u\i--type components of the standard four--body problem.
The occurrence of six extra components is explained by the fact that six
different $(N-2)$--cluster partitions (namely, all the I-- and II--type
partitions) have one additional component for the absorption channel.

The same strategy which led us from Eqs.~(\ref{LS-(N-n+1)}) and
(\ref{fundamental-n}) to Eq.~(\ref{ags-(N-n+1)}) yields
\begin{equation}
{\sf U}^{(N-1)}_{a'b'} =\bar\delta_{a'b'}{{\sf G}_0^{(N-1)}}^{-1}+
\sum_{c'}\bar\delta_{a'c'}{\sf t}^{(N-1)}_{c'}
{{\sf G}_0^{(N-1)}} {\sf U}^{(N-1)}_{c'b'}. \label{newygs}
\end{equation}

      By construction, this result formally has the same structure as
a standard four--body YGS equation. However, the
nature of this equation is different, in that it has extra components
with respect to the standard $N$--body problem, and couples together two
sectors with different numbers of particles.

      For $N=4$ Eq.~(\ref{newygs}) represents our final result. We show
below that the kernel of such an equation is connected after iteration.

For $N>4$ Eq.~(\ref{newygs}) still contains disconnected pieces. Again,
one can proceed in close analogy to the standard GS theory, by recasting
Eq.~(\ref{newygs}) into a formal LS equation:
\begin{equation}
{\sf T}^{(N-2)} ={\sf V}^{(N-2)} +
{\sf V}^{(N-2)} {\sf G}_{0}^{(N-2)} {\sf T}^{(N-2)}.
\label{(N-2)-LS}
\end{equation}
This LS equation operates at a higher hierarchic level, namely
in the standard chain--labelled space for the $N$--body sector,
and in a new space labelled by hybrid chains
for the $(N-1)$--nucleon sector.
The two spaces are spanned by the chain indices $(a' a)$ [ $a\subset a'$],
and $(a',a_1)$ [ $a_1\subset a'$], respectively.  The
hybrid nature of the second chain is obvious, since $a'$ is a
$(N-2)$--cluster partition in the $N$--body sector, and $a_1$ is another
$(N-2)$--cluster partition in the $(N-1)$--nucleon sector.

At this level of the hierarchy, ``Green's functions", ``potentials",
and ``T matrices'' are defined as
\begin{eqnarray}
{{\sf G}_{0}}^{(N-2)}
\equiv&
\left|
\begin{array}{cc}
{G_{0}} ^{(N-2)}_{(a'a|b'b)} & {G_{0}} ^{(N-2)}_{(a'a|b'b_1)} \cr
 {G_{0}} ^{(N-2)}_{(a'a_1|b'b)} & {G_{0}} ^{(N-2)}_{(a'a_1|b'b_1)} \cr
\end{array} \right|
&=
\left| \begin{array}{ccc}
G_0t_aG_0(u_{a'})_{ab}G_0t_bG_0\delta_{a'b'} &\
&
G_0t_aG_0(u_{a'})_{a}g_0\delta_{a'b'} \\
g_0(u^\dagger_{b'})_{b} G_0t_bG_0\delta_{a'b'}
&\ & g_0 (u_{a'}) g_0 \delta_{a'b'}\\
\end{array} \right|,
\nonumber\\ \label{unapalla} \\
{{\sf V}}^{(N-2)}
\equiv & \left|
\begin{array}{cc}
{V} ^{(N-2)}_{(a'a|b'b)} & {V} ^{(N-2)}_{(a'a|b'b_1)} \cr
{V} ^{(N-2)}_{(a'a_1|b'b)} & {V} ^{(N-2)}_{(a'a_1|b'b_1)} \cr
\end{array} \right|
&=
\left| \begin{array}{ccc}
{(G_0t_aG_0)}^{-1}\delta_{ab}\bar\delta_{a'b'} &\ & 0\\
0 &\ &g_0^{-1}\bar\delta_{a'b'} \\
\end{array} \right|,
\label{duepalle}\\
{{\sf T}}^{(N-2)}
\equiv
& \left|
\begin{array}{cc}
{T} ^{(N-2)}_{(a'a|b'b)} & {T} ^{(N-2)}_{(a'a|b'b_1)} \cr
{T} ^{(N-2)}_{(a'a_1|b'b)} & {T} ^{(N-2)}_{(a'a_1|b'b_1)} \cr
\end{array} \right|
&=
\left| \begin{array}{ccc}
U_{a'ab'b} &\ & U_{a'ab'b_1}\\
U^\dagger_{a'a_1b'b} &\ & U_{a'a_1b'b_1} \\
\end{array} \right|.
\end{eqnarray}

It must be clearly understood that {\it when $a'$ and $b'$ are of type
{\rm III} or {\rm IV} all the hybrid--chain components are absent},
whereas these hybrid--chain components are substantial when $a'$, $b'$
are of type I and II. As stated before, for any given I-- or II--type
partition $a'$, there is one and only one partition $a_1$ in the
($N$-1)--nucleon sector; hence, $\delta_{a'b'}$ $\Rightarrow$
$\delta_{a_1b_1}$, as well as $\bar\delta_{a_1b_1}$
$\Rightarrow$ $\bar\delta_{a'b'}$. To make these remarks clearer, it is
instructive to write Eq.~(\ref{(N-2)-LS}) [or (\ref{newygs})]
explicitly. One has
\begin{mathletters}
\begin{eqnarray}
U_{a'a b'b} &=&
\bar\delta_{a'b'}\delta_{ab}(G_0t_aG_0)^{-1} +
{\displaystyle \sum_{c'c}}
\bar\delta_{a'c'}(u_{c'})_{ac}G_0t_cG_0U_{c'c b'b}\nonumber\\
& &+{\displaystyle \sum_{c'(\in {\rm I, II})}}\bar\delta_{a'c'}(u_{c'})_ag_0
U^{\dagger}_{c'c_1b'b} ,\label{casinoa}\\
U^{\dagger}_{a'a_1 b'b} &=&
{\displaystyle \sum_{c'(\in {\rm I, II})c}}
\bar\delta_{a'c'}(u^{\dagger}_{c'})_cG_0t_cG_0
U_{c'c b'b} + {\displaystyle \sum_{c'(\in {\rm I, II})}}
\bar\delta_{a'c'}u_{c'}g_0U^{\dagger}_{c'c_1b'b} ,\label{casinob}\\
U_{a'a b'b_1} &=& {\displaystyle \sum_{c'c}}
\bar\delta_{a'c'}(u_{c'})_{ac}G_0t_cG_0 U_{c'c b'b_1} +
{\displaystyle \sum_{c'(\in {\rm I, II})}}\bar\delta_{a'c'}(u_{c'})_ag_0
U_{c'c_1b'b_1} ,\label{casinoc}\\
U_{a'a_1 b'b_1} &=& g_0^{-1} \bar\delta_{a'b'} +
{\displaystyle \sum_{c'(\in {\rm I, II})c}}
\bar\delta_{a'c'}(u^{\dagger}_{c'})_cG_0t_cG_0U_{c'c b'b_1}\nonumber\\
& & + {\displaystyle \sum_{c'(\in {\rm I, II})}}\bar\delta_{a'c'}u_{c'}g_0
U_{c'c_1b'b_1} .\label{casinod}
\end{eqnarray}
\end{mathletters}
      As a consequence of the use of the fundamental ansatz
(\ref{newfundamental}) the transition operators $U_{ab}$,
$U^{\dagger}_a$, and $U_a$, labelled by partition indices, are
now replaced by chain--labelled operators. The Eqs.~(\ref{casinoa})
and (\ref{casinob}) couple operators referring to $\pi$ scattering
and absorption, whereas pion production is involved in the second
set [Eqs.~(\ref{casinoc}) and (\ref{casinod})].

\subsection{Connectedness properties for the $\pi$--trinucleon case}

      The well--known results about the connectedness of the FY kernel
cannot be extended straightforwardly to the kernel of Eq.~(\ref{(N-2)-LS}),
even in the $\pi$NNN--NNN case. Indeed, we have seen that for $N=4$
Eq.~(\ref{(N-2)-LS}) couples 24 chain--labelled components, instead of
the 18 components of the standard four--body problem. Moreover, the
vertex operators act differently from the two--body interactions in
determining the topological structure of the graphs associated to the
perturbative expansion of the kernel. This is why we devote
this subsection to a detailed discussion of connectedness for the
$\pi$NNN--NNN problem. We shall show that the kernel of Eq.~(\ref{(N-2)-LS})
is actually connected after three iterations.
\par
      Let us consider the lowest--order approximation ${\sf K}$ to the
$\pi$NNN--NNN kernel, in which the subamplitudes appearing in the
``Green's function" ${\sf G}^{(2)}_0$ are replaced by the driving terms of
the corresponding dynamical equations (\ref{III-AB}),
(\ref{I-ABpp})--(\ref{I-AB}) or  (\ref{II-ABpp})--(\ref{II-AB}). One has
\begin{equation}
{\sf K} = \left|
\begin{array}{cc}
K_{(a'a|b'b)} & K_{(a'a|b'b_1)} \\
K_{(a'a_1|b'b)} & K_{(a'a_1|b'b_1)} \\
\end{array} \right|
= \left|
\begin{array}{cc}
\bar\delta_{a'b'}\bar\delta_{ab}\delta_{a\subset b'} t_bG_0 &
\quad \bar\delta_{a'b'}(f_{b'})_a g_0 \\
\bar\delta_{a'b'}(f_{b'}^{\dagger})_b G_0t_bG_0 &
\quad \bar\delta_{a'b'}{\cal V}_{b'}g_0 \\
\end{array} \right|
\equiv \left|
\begin{array}{cc}
K_{(1|1)} &\ K_{(1|0)} \\
K_{(0|1)} &\ K_{(0|0)} \\
\end{array} \right| .\label{ker}
\end{equation}
Clearly if ${\sf K}$ can be shown to be connected after iteration,
the same is true for the full kernel.

      In Eq.~(\ref{ker}) we have introduced a simplified notation,
where the actual dependence of the kernel elements upon the
chain indices is omitted, and the presence or absence of the pion
is emphasized. Thus, $K_{(1|0)}$ is associated to an elementary
process in which the pion is produced through the vertex $(f_{b'})_a$.
This notation will turn out to be efficient and useful when discussing
the various iterations of ${\sf K}$.
\par
      In view of the non--standard structure of ${\sf K}$, we
follow a step--wise procedure, {\it i.e.}, we isolate the
{\it connected} terms of ${\sf K}^2$ first, then we proceed to
identify the connected pieces of ${\sf K}^3$, and so on, until
full connectedness (for ${\sf K}^4$) is exhibited. Before embarking
on the discussion of connectedness, however, it is convenient to
examine the structure of the internal vertices $(f_{a'})_a$ and
$(f_{a'}^{\dagger})_a$ more carefully. For the sake of simplicity,
we limit ourselves to the production vertex, the same considerations
obviously applying to the absorption operator.
If $a'$ is a I--type partition [$a' = ((\pi ij)k)$], we have two
possibilities, either $a$ is the NN pair $(ij)$, or it is a $\pi$N
pair [$a = i$, say]. In the former case one gets from Eq.~(\ref{intvert})
$(f_{a'})_a = f(i) + f(j)$, {\it i.e.} the pion can be produced by
{\it both} nucleons in the pair $a$. This is graphically depicted in
Fig.\ \ref{vertex1}. In the latter case, one has $(f_{a'})_a = f(j)$,
namely, the pion is produced by the nucleon internal to $a'$ which is
{\it not} contained in $a$ (see Fig.\ \ref{vertex2}). As for the II--type
partitions [$a' = ((\pi j) (ik))$], the internal vertex can be different from
zero {\it only if} $a$ is the two--nucleon pair $(ik)$, in which case one
gets $(f_{a'})_a = f(j)$ again, as it was to be expected on physical
grounds.

After the first iteration one gets
\begin{equation}
{\sf K}^{2} = \left|
\begin{array}{cc}
K_{(1|1)} K_{(1|1)} + K_{(1|0)} K_{(0|1)}&\
K_{(1|1)} K_{(1|0)} + K_{(1|0)} K_{(0|0)}\cr
K_{(0|1)} K_{(1|1)} + K_{(0|0)} K_{(0|1)}&\
K_{(0|1)} K_{(1|0)} + K_{(0|0)} K_{(0|0)}\cr
\end{array} \right| \equiv \left|
\begin{array}{cc}
K^2_{(1|1)} &\  K^2_{(1|0)} \\
K^2_{(0|1)} &\  K^2_{(0|0)} \\
\end{array} \right| . \label{ker2}
\end{equation}

      The explicit dependence of ${\sf K}^2$ upon the chain labels
can be easily exhibited, once Eq.~(\ref{ker}) is taken into account.
For instance, for $K_{(0|1)}K_{(1|1)}$ one has
\begin{equation}
K_{(0|1)}K_{(1|1)} = \sum_{cc'}\bar\delta_{a'c'} (f^{\dagger}_{c'})_c
G_0 t_c G_0 \bar\delta_{c'b'} \bar\delta_{cb}\delta_{c\subset b'}
t_b G_0 ,\label{k201}
\end{equation}
whereas for $K^2_{(0|0)}$ one gets
\begin{equation}
K^{2}_{(0|0)} = \sum_{cc'}\bar\delta_{a'c'} (f^{\dagger}_{c'})_c
G_0 t_c G_0 \bar\delta_{c'b'} (f_{b'})_c g_0
+ \sum_{c'}\bar\delta_{a'c'}{\cal V}_{c'} g_0\bar\delta_{c'b'}
{\cal V}_{b'} g_0 .\label{k200}
\end{equation}

      Let us consider the structure of this term
more closely. The second contribution in Eq.~(\ref{k200}) occurs only
if both $b'$ and $c'$ refer to I--type partitions and, because of the
$\bar\delta$--conditions, is {\it already connected}. Actually, this term
has the same structure as the iteration of the standard Faddeev kernel.
The connectedness of the former term in
Eq.~(\ref{k200}) is less trivial, owing to the presence of the internal
vertices, through which the NNN space communicates with the $\pi$NNN
sector. As we observed before, both $b'$ and $c'$ must be of the
I or II type. Let us suppose that both of them are of type I, so that
$b' = ((\pi ij)k)$ and $c' = ((\pi ik)j)$ (we recall that $c' \neq b'$).
Since $c\subset c', b'$, one necessarily has $c = i$, $(f_{b'})_c = f(j)$
and $(f_{c'}^{\dagger})_c = f(k)$, so that the corresponding graph is
connected (see Fig.\ \ref{grak2oo}). By a similar reasoning one proves that
the former contribution in (\ref{k200}) is connected for $b'$ of type I and
$c'$ of type II, or viceversa. A typical contribution is shown in
Fig.\ \ref{grak2oop}. Finally, for both $b'$ and $c'$ of type II, one
cannot find a common pair $c$, so that for this case there is
no contribution to $K_{(0|1)}K_{(1|0)}$. It is worth to observe that
these terms provide {\it three--nucleon forces} in the NNN space, due to
intermediate $\pi$ exchange in the full, four--body sector.

      Also the term $K_{(0|1)}K_{(1|1)}$ is free of dangerous, disconnected
graphs. The analysis is somewhat more lengthy than before, since $b'$
can be of any type now (I, II or III; we recall that there are no IV--type
partitions for $N$=4). However, the intermediate--partition index
$c'$ in Eq.~(\ref{k201}) is limited to the I or II class only. Again,
one cannot have both $b'$ and $c'$ of type II [with $b' \neq c'$], because
there is no pair $c$ common to both. One can verify by inspection that all
other possibilities lead to connected contributions, typical examples
being illustrated in Figs.\ \ref{grak2o1} and\ \ref{grak2o1p}. On the
other hand, disconnected contributions do arise in the other terms
of ${\sf K}^2$, so that one has to consider also the cube of the kernel
${\sf K}$
\begin{equation}
{\sf K^3} \equiv \left|
\begin{array}{cc}
K^3_{(1|1)} &\quad K^3_{(1|0)} \\
K^3_{(0|1)} &\quad K^3_{(0|0)} \\
\end{array} \right| \label{kcubo}
\end{equation}
where
\begin{mathletters}
\begin{eqnarray}
K^3_{(1|1)} =&
K_{(1|1)}K_{(1|1)}K_{(1|1)} +
K_{(1|0)}K_{(0|1)}K_{(1|1)} +&
K_{(1|1)}K_{(1|0)}K_{(0|1)} \nonumber\\
\ &\ &+ K_{(1|0)}K_{(0|0)}K_{(0|1)} ,\label{k311}\\
K^3_{(1|0)} =&
K_{(1|1)}K_{(1|1)}K_{(1|0)} +
K_{(1|1)}K_{(1|0)}K_{(0|0)} +&
K_{(1|0)}K^2_{(0|0)} ,\label{k310}\\
K^3_{(0|1)} =&
K_{(0|1)}K_{(1|1)}K_{(1|1)} +
K_{(0|0)}K_{(0|1)}K_{(1|1)} +&
K^2_{(0|0)}K_{(0|1)} ,\label{k301}\\
K^3_{(0|0)} =&
K_{(0|1)}K_{(1|1)}K_{(1|0)} +
K_{(0|0)}K_{(0|1)}K_{(1|0)} +&
K^2_{(0|0)}K_{(0|0)} .\label{k300}
\end{eqnarray}
\end{mathletters}
The second term of $K^3_{(1|1)}$, the first two terms of $K^3_{(0|1)}$
and the first term of $K^3_{(0|0)}$ are clearly connected, because
they contain the connected factor $K_{(0|1)}K_{(1|1)}$. Similarly, the
last terms of $K^3_{(1|0)}$, $K^3_{(0|1)}$ and $K^3_{(0|0)}$ are
connected because of the presence of the factor $K^2_{(0|0)}$.
Finally, the second contribution to $K^3_{(0|0)}$ is connected
by the factor $K_{(0|1)}K_{(1|0)}$. We are left with the first,
third, and last term in $K^3_{(1|1)}$, and with the first two terms
of $K^3_{(1|0)}$.

      The term $K_{(1|1)}K_{(1|1)}K_{(1|1)}$ in Eq.~(\ref{k311})
is nothing else but the second iteration of a {\it standard
Faddeev--Yakubovski\u\i\ kernel}, and is therefore certainly connected.
Let us consider the term $K_{(1|1)}K_{(1|0)}K_{(0|1)}$. In this
case one cannot resort to previous results, so that the various
contributions to this matrix element have to be checked by inspection
so as to ascertain the possible presence of disconnected graphs.
In virtue of Eq.~(\ref{ker}) one has
\begin{equation}
K_{(1|1)}K_{(1|0)}K_{(0|1)} = \sum_{cc'}\sum_{d'}
\bar\delta_{a'c'}\bar\delta_{ac}\delta_{a\subset c'}t_cG_0
\bar\delta_{c'd'}(f_{d'})_c g_0
\bar\delta_{d'b'}(f_{b'}^{\dagger})_b G_0t_bG_0 \ .\label{kkk}
\end{equation}
With reference to $b'$ and $d'$ one has four possibilities, since both
$b'$ and $d'$ can be I--type or II--type partitions (partitions of class
III are excluded because of the presence of the vertex functions). If
$b' = ((\pi ij)k)$ and $d' = ((\pi jk)i)$ [$b' \neq d'$], one could
have, for instance, a first scattering between the pion and nucleon $i$
[$b=i$]. Then the pion can be only absorbed by nucleon $j$. One has
now two possibilities, either $t_c$ describes a $\pi$N interaction
[$c=k$, for instance], or it describes an NN rescattering [$c=(jk)$].
In both cases the structure of the production vertex ensures that
the corresponding graph is connected [see Fig.\ \ref{3ord}]. We repeated this
analysis for the various contributions to Eq.~(\ref{kkk}), and found no
disconnected graphs. Similarly, connected graphs only have been found
for $K_{(1|1)}K_{(1|1)}K_{(1|0)}$ in $K^3_{(1|0)}$. The situation is
different for the remaining terms in Eqs.~(\ref{k311}) and (\ref{k310}).
Here, disconnected graphs occur, describing the successive rescatterings
between a couple of nucleons, the pion being produced and/or absorbed
by the spectator nucleon [see Fig.\ \ref{disco}]. One is therefore forced
to go to the next iteration of ${\sf K}$, in order to analyze the terms
arising from $K_{(1|0)}K_{(0|0)}K_{(0|1)}$ and
$K_{(1|1)}K_{(1|0)}K_{(0|0)}$. As for the former term, after iteration
it yields the two contributions
\[K_{(1|0)}K_{(0|0)}K_{(0|1)}K_{(1|1)}\quad {\rm and} \quad
K_{(1|0)}K_{(0|0)}K_{(0|1)}K_{(1|0)}\]
to ${\sf K}^4$. They are both connected owing to the presence of the
factors $K_{(0|1)}K_{(1|1)}$ and $K_{(0|1)}K_{(1|0)}$, respectively.
Let us consider the latter, third--order disconnected contribution.
After iteration it gives the fourth--order terms
\[K_{(1|1)}K_{(1|0)}K_{(0|0)}K_{(0|1)} \quad {\rm and} \quad
K_{(1|1)}K_{(1|0)}K_{(0|0)}K_{(0|0)} .\]
      The latter of these matrix elements is connected because of the
presence of the factor $K_{(0|0)}K_{(0|0)}$. The former requires
some more thought. It can be regarded as the amplitude
$K_{(1|0)}K_{(0|0)}K_{(0|1)}$ [described by Fig.\ \ref{disco}] times the
kernel element $K_{(1|1)}$. This factor ensures that a two--body
t--matrix connects the pair of nucleons emitting the pion, thereby
rendering the full graph connected [see Fig.\ \ref{4order}].
\par
      In conclusion, a careful analysis of the kernel ${\sf K}$
of the $\pi$NNN--NNN equations (\ref{(N-2)-LS}) and of its
iterations reveals no pathological (disconnected) terms after the
third iteration. This result, and the peculiar way in which
connectedness is here achieved have to be contrasted with what happens
in the standard four--body problem, where connectedness is
achieved after two iterations. This is due to the further disconnected
structures introduced by the possible creation or destruction of
the pion.

\section{Summary}

      In this paper we have introduced a new approach to the
$\pi$--multinucleon problem. The main results of the paper can
be summarized as follows.
\par
      We have first solved the problem of the subsystem
dynamics, and then, through a suitable generalization of the
standard chain--of--partitions--labelled formalism, we have
derived the coupled integral equations for the consistent and
simultaneous treatment of the dynamics in the no--pion and
one--pion sectors.
\par
      For the $\pi$--trinucleon system, we have shown that
the kernel of such equations is connected after three iterations.
In so doing, we have analyzed how connectedness is achieved in
the various sectors of the theory.
\par
      To obtain this generalization, in the first part of the paper
we have presented the standard $N$--body theory under a new, suitable
perspective.

\figure{Graphical representation of the production vertex $(f_{a'})_a$
for $a'=((\pi ij)k)$ and $a=(ij)$. The full lines represent the
nucleons, the dashed one is associated to the pion.\label{vertex1}}
\figure{Graphical representation of the production vertex $(f_{a'})_a$
for $a'=((\pi ij)k)$, $a=i$, or for $a'=((\pi j)(ik))$,
$a=(ik)$.\label{vertex2}}
\figure{A connected contribution to the first term in $K^2_{(0|0)}$
with $b'=((\pi ij)k)$ and $c'=((\pi ik)j)$. The blob represents a
two--body t--matrix.\label{grak2oo}}
\figure{A connected contribution to the first term in $K^2_{(0|0)}$
with $b'=((\pi ij)k)$ and $c'=((\pi k)(ij))$.\label{grak2oop}}
\figure{A connected contribution to the first term in $K^2_{(0|1)}$
for $b'=((ijk)\pi)$ and $c'=((\pi i)(jk))$. Here, $b=(ij)$,
$c=(jk)$.\label{grak2o1}}
\figure{A connected contribution to the first term in $K^2_{(0|1)}$
for $b'=((\pi i)(jk))$ and $c'=((\pi jk)i)$. Here, $b=i$,
$c=(jk)$.\label{grak2o1p}}
\figure{Connected contributions to $K_{(1|1)}K_{(1|0)}K_{(0|1)}$
with $b'=((\pi ij)k)$, $b=i$ and $d'=((\pi jk)i)$. $a$) For
$c=k$, $b$) for $c=(jk)$.\label{3ord}}
\figure{A disconnected contribution to $K_{(1|0)}K_{(0|0)}K_{(0|1)}$.
The wavy line represents the NN potential in the three--nucleon
space.\label{disco}}
\figure{A connected contribution to the fourth--order term
$K_{(1|1)}K_{(1|0)}K_{(0|0)}K_{(0|1)}$.\label{4order}}


\begin{references}
\bibitem{note2} The expression (\ref{eq:ampnp}) for the
$2\leftarrow 2$ transition amplitudes has been given, in a slightly
different form, in Ref.~\cite{sa80}. It has been derived through a
different, more lengthy procedure, in the second of Refs.~\cite{cv83f}.
\end{references}
\end{document}